\documentclass[letterpaper, 10pt, conference]{ieeeconf}      

\IEEEoverridecommandlockouts                              

\usepackage{graphics} 
\usepackage{mathptmx} 
\usepackage{times} 
\usepackage{bbm}
\usepackage{bm}
\usepackage{amsmath} 
\usepackage{amssymb}  
\usepackage{mathrsfs}
\usepackage[ruled,vlined,linesnumbered]{algorithm2e}
\newtheorem{remark}{Remark}

\newtheorem{lemma}{Lemma}
\usepackage{mathtools}
\usepackage{eucal}
\usepackage{url}
\graphicspath{{Figures/}}
\renewcommand{\baselinestretch}{.94}

\title{\LARGE \bf
Autonomous UAV Navigation in an Unknown Environment via Multi-Trajectory Model Predictive Control
}

\author{Danilo Saccani and Lorenzo Fagiano
	\thanks{This research has been supported by the Italian Ministry of University and Research (MIUR) under the PRIN 2017 grant n. 201732RS94 ``Systems of Tethered Multicopters''.}
\thanks{The authors are with the Dipartimento di Elettronica, Informazione e Bioingegneria (DEIB), Politecnico di Milano, Italy. {\tt\small \{danilo.saccani, lorenzo.fagiano\}@polimi.it}}%
}

\begin{document}

\maketitle
\thispagestyle{empty}
\pagestyle{empty}

\begin{abstract}
A novel model predictive control (MPC) formulation, named multi-trajectory MPC (mt-MPC), 
is presented and applied to the problem of autonomous navigation of an unmanned aerial vehicle (UAV) in an unknown environment.
The UAV is equipped with a LiDAR sensor, providing only a partial description of the surroundings and resulting in time-varying constraints as the vehicle navigates among the  obstacles. The control system layout is hierarchical: the low-level loops stabilize the vehicle's trajectories and track the set-points commanded by the high-level, mt-MPC controller. The latter is required to plan the UAV trajectory trading off safety, i.e. to avoid collisions with the uncertain obstacles, and exploitation, i.e. to reach an assigned target location.
To achieve this goal, mt-MPC considers different future state trajectories in the same Finite Horizon Optimal Control Problem (FHOCP), enabling a partial decoupling between constraint satisfaction (safety) and cost function minimization (exploitation). Recursive feasibility and, consequently, persistent obstacle avoidance guarantees are derived under the assumption of a time invariant environment.
The performance of the approach is studied in simulation and compared with that of a standard MPC, showing good improvement.
\end{abstract}
\section{Introduction}\label{S:introduction}
In the last decade, autonomous vehicles have received broad attention thanks to their high impact on many areas of science, technology, and society \cite{bagloee2016autonomous}, \cite{chung2018survey}. 
We consider here autonomous unmanned aerial vehicles (UAVs), or drones. A key field of research is the study of autonomous navigation strategies to reach a desired target in presence of obstacles. 
This problem entails different challenges \cite{tang2018autonomous}, such as the need to plan a feasible trajectory, satisfying actuator constraints and avoiding collisions, to retain the  
real-time applicability of the algorithm.
When the obstacles' position and shape are perfectly known, several approaches exist in the literature for off-line path planning, such as potential field
, traditional search (e.g., A*), 
sampling-based planners 
or nonlinear trajectory generation 
\cite{tang2018autonomous}.
The use of model predictive control (MPC) has been investigated as well, in view of the possibility to cope with multiple-inputs, multiple-outputs systems and to manage  constraints \cite{BemporadMILP,oleynikova2016continuous}. In \cite{BemporadMILP,richards2002aircraft}, the control problem to compute the optimal trajectory is formulated as a mixed-integer linear program, where an integer variable checks the belonging of the optimal trajectory to the free space.
Solutions where no integer variables are required can be found in \cite{BemporadLTVconstr} and \cite{mousavi2013ltv}, where a linear time-variant (LTV) MPC strategy is used as an high-level navigation controller and the free space is under-approximated with convex polytopes, still assuming that position and shape of the surrounding obstacles are perfectly known. 
\begin{figure}
	\centering
	\includegraphics[width=0.8\columnwidth]{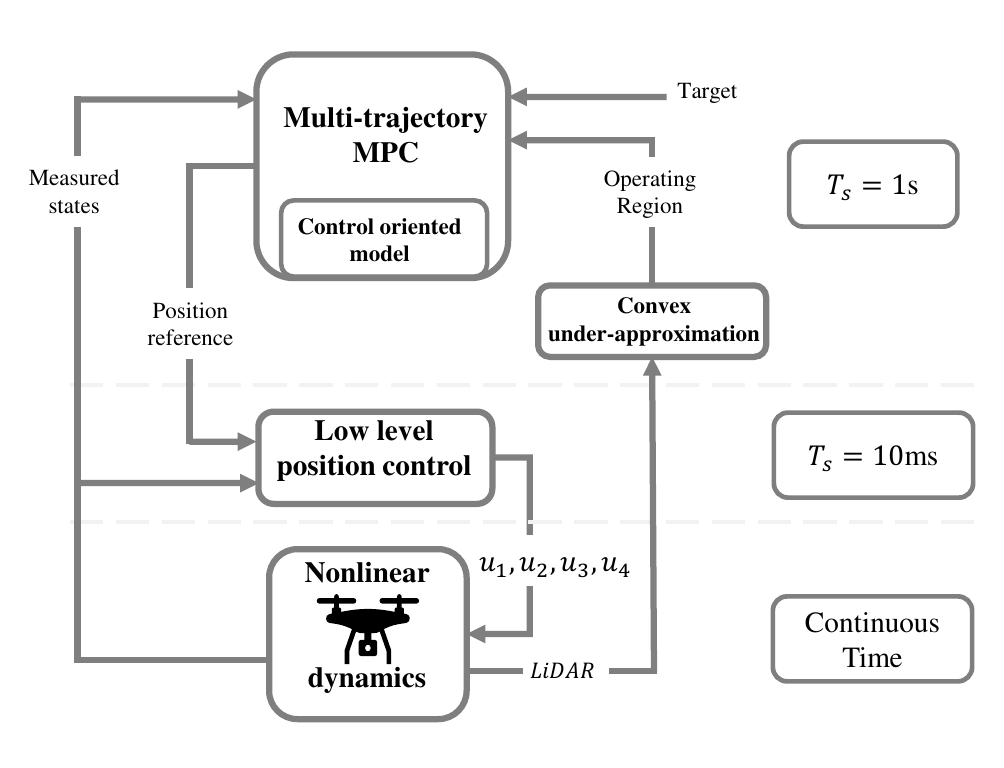}
	\caption{Hierarchical structure of the UAV control system: the low-level position controller, described in  \cite{FagianoSTEM}, tracks the reference signals provided by a high-level navigation controller. The latter elaborates the LiDAR readings (Section \ref{SS:LiDAR}) to build a convex under-approximation of the feasible space (Section \ref{SS:convex-approx}), which is employed, together with a simplified, control-oriented model of the drone (Section \ref{SS:control-model}), by the state-feedback, multi-trajectory MPC strategy (Section \ref{SS:mtMPC}). 
		\label{fig:hierarchical}}
\end{figure}
When, instead, the environment is unknown, the motion planning can only use partial information collected by the available sensors to generate a feasible path, and it can not be carried out off-line. 
In this case, the control  logic has to balance two conflicting aspects: \textit{safety}, that is to avoid obstacles, and \textit{exploitation}, i.e. to reach the desired target in short time. Proposed approaches to deal with this problem include 
Reinforcement Learning (RL) methods (e.g. \cite{sharma2012autonomous}), 
which employ directly the sensor measurements, or combinations of Optimal Control and Learning, as presented in \cite{greatwood2019reinforcement}. 
In this paper, we propose a new approach to obtain safe navigation to the target in an a priori unknown environment, considering a multi-copter drone equipped with a planar LiDAR (Light Detection And Ranging) sensor and featuring a hierarchical control structure: at a low level, standard control loops stabilize the vehicle's trajectories and track the set-points provided by the high-level navigation logic. The latter is designed with a novel approach for optimal constrained control, named multi-trajectory MPC (mt-MPC), able to trade-off the mentioned conflicting aspects in an intuitive way. Specifically, mt-MPC considers different future state trajectories in the same Finite Horizon Optimal Control Problem (FHOCP), enabling a partial decoupling between constraint satisfaction (safety) and cost function minimization (exploitation). 
In addition to introducing the mt-MPC concept, we deal with two application-specific problems: the need to derive a convex under-approximation of the free space around the drone, in order to formulate the FHOCP as a convex quadratic program (QP) which is solved efficiently, and the need to navigate around obstacles that stand between the drone and its target. We solve the first problem with a new method that exploits the LiDAR readings, and the second one through a temporary target shifting procedure. A consequence of dealing with an unknown environment is the presence of time-varying constraints, such that standard receding horizon strategies do not guarantee recursive feasibility anymore. To solve this problem, we propose a modified receding horizon implementation and prove recursive feasibility, hence persistent obstacle avoidance, under the assumption of time invariant environment. 
We eventually test the approach in simulation with a  realistic 6-degrees-of-freedom drone model, and compare the obtained performance with that of a ``standard'' (i.e., single trajectory) MPC.
\section{System and environment models, problem formulation}
\label{S:systModel and problem formulation}
The three main ingredients defining the problem at hand are: the autonomous system, the LiDAR sensor, and the environment. 
The autonomous quad-copter features a hierarchical control structure, with a high-level planner that computes the reference $\bm{P_{ref}}=[P^{x_f}_{ref},\,P^{y_f}_{ref},\,P^{z_f}_{ref}]^T$ for the low-level position controller, see Fig. \ref{fig:hierarchical} (bold-faced symbols denote vectors in the 3D space). 
\\
The latter, similar to the one employed in \cite{FagianoSTEM}, is not described here for the sake of conciseness. A different low-level controller can be employed as well, as long as it is able to stabilize the drone's trajectory and to track the given reference position. The low-level controller is static and it is implemented at a high-enough sampling frequency (typically 50-100 Hz) to obtain good positioning performance. 
\subsection{Control-oriented model}\label{SS:control-model}
\begin{figure}[!htb]
	\centering
	\includegraphics[width=0.8\columnwidth]{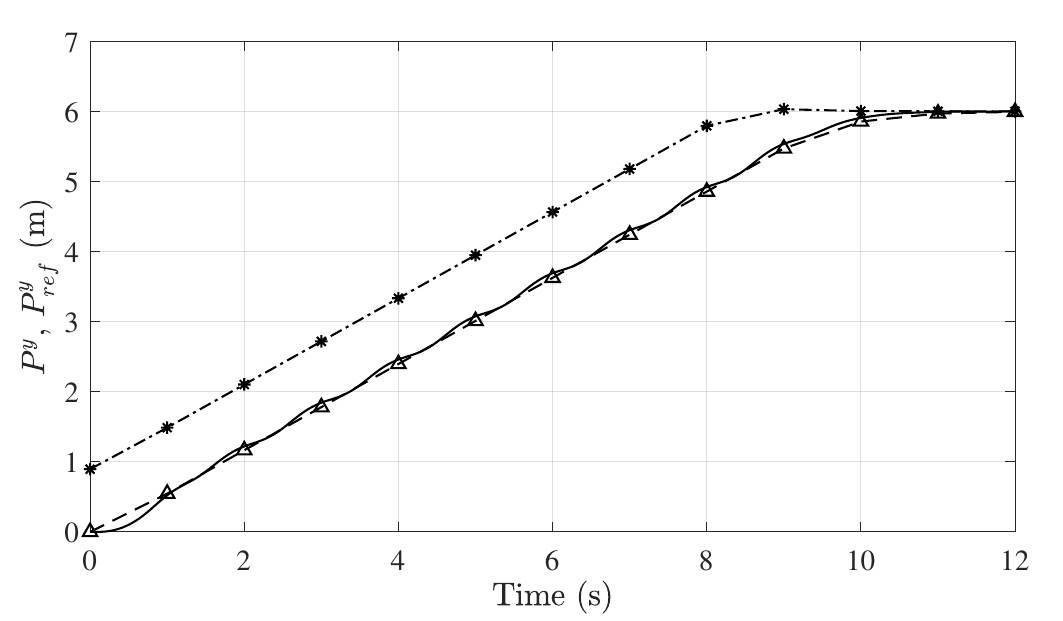}
	\caption{Example of $P^{y_f}$ response of the nonlinear quad-copter model  
		(solid line) to a  $P^y_{ref}$ reference ramp (dash-dotted line with `*'), compared with that of the control-oriented model \eqref{eq:control-oriented} (dashed line with `$\vartriangle$'). Similar behavior is obtained for the other position components.} 
	\label{fig:ControlOrientModel}
\end{figure}

Our high-level control strategy (see Fig. \ref{fig:hierarchical}) employs a linear time invariant (LTI) model of the drone in close loop with the low-level controller, named ``control-oriented'' model. An LTI model structure is suitable to describe the  dynamics of the position-controlled real drone, moreover it leads to a convex QP to be solved at each time $k$. To obtain an accurate control-oriented model of the drone dynamics,  
we first consider the following continuous time double integrator with state feedback:
\begin{equation}\label{eq:ct-system}
\begin{aligned}
\begin{bmatrix}
\dot{\bm{P}}(t)\\
\dot{\bm{V}}(t)
\end{bmatrix}
= \begin{bmatrix}
0^{3\times3} & \mathbb{I}^{3\times3} \\
-K_1 & -K_2
\end{bmatrix}
\begin{bmatrix}
\bm{P}(t)\\
\bm{V}(t)
\end{bmatrix}
+
\begin{bmatrix}
0^{3\times3}\\
K_1
\end{bmatrix}
\bm{P}_{ref}(t)
\end{aligned}
\end{equation}
where $\bm{P}(t)\in \mathbb{R}^3$, $\bm{V}(t)\in \mathbb{R}^3$ are the drone position and velocity, and we denote with 
\begin{equation}\label{eq:acceleration}
\bm{A}(t)=K_1\left(\bm{P}_{ref}(t)-\bm{P}(t)\right)-K_2\bm{V}(t)
\end{equation} its acceleration vector. In \eqref{eq:ct-system}, $0^{a\times b}$ and $\mathbb{I}^{a\times b}$ denote respectively a matrix of zeros and the identity matrix of size ${a\times b}$. $K_1\in \mathbb{R}^{3\times3}$, $K_2\in \mathbb{R}^{3\times3}$ are suitable gains that we tuned by carrying out a system identification procedure to obtain a closed loop system response as close as possible to the one of the actual nonlinear system as shown in Fig. \ref{fig:ControlOrientModel}. We then convert \eqref{eq:ct-system} to discrete-time with sampling time $T_s$, obtaining the wanted control-oriented model of the form
\begin{equation}\label{eq:control-oriented}
x(k+1)=Ax(k)+B\bm{u}(k),
\end{equation}
where $x(k)=[P^{x_f}(k),P^{y_f}(k),P^{z_f}(k),V^{x_f}(k),V^{y_f}(k),V^{z_f}(k)]^T \in \mathbb{R}^6$ is the state of the position-controlled system and $\bm{u}(k)=[P^{x_f}_{ref}(k),P^{y_f}_{ref}(k),P^{z_f}_{ref}(k)]^T\in \mathbb{R}^3$ its input. Thus, the LTI model \eqref{eq:control-oriented} features $\bm{P_{ref}}(k)$ as input, consistently with our setup, and its state includes both position and velocity, making it possible to easily include in the predictive control strategy constraints on these quantities and on the accelerations as well, through the linear equation \eqref{eq:acceleration}. 

\subsection{LiDAR sensor}\label{SS:LiDAR}
We assume that the drone is equipped with a 360-degrees planar LiDAR sensor that continuously scans the surrounding environment on an horizontal plane (i.e., parallel to $(x_f,y_f)$).
The LiDAR is installed on an active gimbal able to maintain the sensor horizontal in spite of vehicle's motion.
Let us denote with $k\in\mathbb{Z}$ the discrete time variable at sampling time $T_s$ employed by the high-level navigation strategy.
The LiDAR provides a vector $\rho(k)\in\mathbb{R}^M$ of $M=\frac{2\pi}{\alpha_s}$ readings, where $\alpha_s$ is the angular resolution.
Each entry $\rho_i(k),\,i=0,\ldots,M-1$ is the measured distance between the sensor and the closest obstacle along the 2D unit vector $\left[\cos{\left(\alpha_i\right)},\,\sin{\left(\alpha_i\right)}\right]^T$, where $\alpha_i=i\,\alpha_s,\,i=0\ldots,M-1$.
Each measurement can thus be expressed as a vector $s_i(k)=\rho_i(k) [\cos{(\alpha_i)},\,\sin{(\alpha_i)}]^T, \, i=0,\dots,M-1$. To simulate the sensor, we use an obstacle detection routine able to check the belonging of a ray to a convex polytope representing the obstacle in the probed direction, similar to the one described in \cite{BOFA20_arxiv}. Note that the use of convex obstacles for this routine doesn't lead to a loss of generality, since non convex shapes can be modeled by partially overlapping convex ones.
Finally, the LiDAR has a maximum detection range $R_L$, such that if no obstacle is detected in the $i$-$th$ direction, then $\rho_i=R_L$.
\subsection{Environment model and problem formulation}\label{SS:problem formulation}
\label{sec:Env}
For simplicity, the environment considered in this work is composed of a number $N_o$ of three-dimensional obstacles with constant cross-section parallel to the $(x_f,y_f)$ plane and an infinite height. 
As a matter of fact, the proposed approach can be extended with little modifications to obstacles with variable cross-section and the use of a 3D LiDAR sensor; such an extension is currently under development. In this framework, each obstacle can be then described as a compact set $O_i\in\mathbb{R}^2,\,i=1,\ldots,N_o$. Let us denote with $v_{(a:b)}$ the entries from the $a$-th to $b$-th position of the generic vector $v$. Now, recalling that the first two entries of vector $x$ of model \eqref{eq:control-oriented} correspond to the $(x_f,y_f)$ position of the drone, we can define the obstacles' set $\mathcal{O}$ as:
\begin{equation}
\mathcal{O}\doteq\left\{x\in\mathbb{R}^6:x_{(1:2)}\in O_i \text{ for some $i=1,\ldots,N_o$ }\right\},
\end{equation} 
\noindent The problem we consider is the design of a high-level, discrete-time navigation logic that exploits the LiDAR measurements $s_i(k), \, i=0,\dots,M-1$ and the state $x(k)$ of model \eqref{eq:control-oriented} to compute, at each step $k$, the reference $\bm{P}_{ref}(k)=\bm{u}(k)$ in order to move the UAV from a given initial state $x(0)\notin\mathcal{O}$ towards a target position $\bm{P}_t(k)\in\mathbb{R}^3$ in a safe way, i.e. such that
$x(k)\notin\mathcal{O},\forall k\geq0$.
\section{Proposed approach}
\label{sec:mtMPC}
Fig. \ref{fig:hierarchical} presents the layout of the proposed high-level strategy. At each time step, the controller elaborates the LiDAR readings to build a convex under-approximation of the free space around the drone. The latter is used to enforce obstacle avoidance in the proposed mt-MPC technique. 

\subsection{Convex approximation of the free space}\label{SS:convex-approx}
\label{subsec:poly}
\begin{figure}[!hbt]
	\centering
	\includegraphics[width=.75\columnwidth]{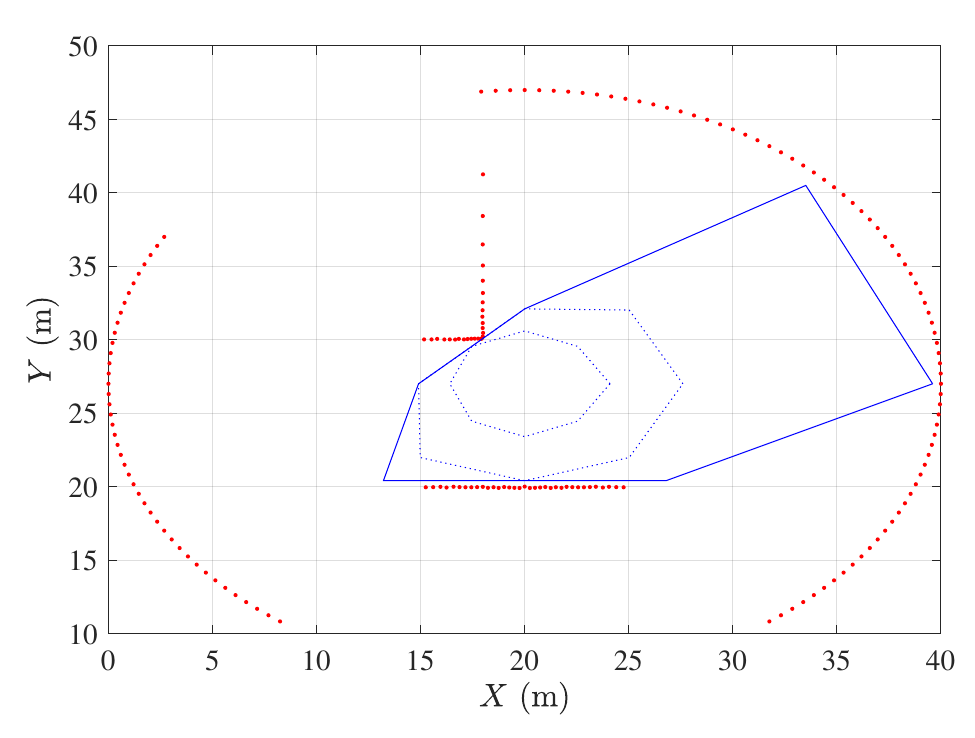}
	\caption{Example of construction of the convex under-approximation of the free region: intermediate steps (dotted blue lines) and final result (solid polytope). LiDAR readings are shown as red points.}	
	\label{fig:build_poly}
\end{figure}

Let us consider the position of the drone at time $k$ projected on the $(x_f,y_f)$ plane, i.e. $x_{(1:2)}(k)$, and let us denote with  $\mathcal{L}(k)\doteq\{s_1(k),\ldots,s_M(k)\}\in\mathbb{R}^2$ the collection of $M$ LiDAR observations at time $k$, which approximate the perimeter of a generally non-convex, obstacle-free region. Given a set $B\in\mathbb{R}^2$, condition $\mathcal{L}(k)\notin B$ means that no one of the LiDAR readings belongs to $B$. Let $d_{step}>0$ be a user-defined distance. Moreover, we consider a user-defined number $n_v$ of equally-spaced fixed directions $\alpha_i^v$ on the plane $(x_f,y_f)$, i.e. $\alpha_{i+1}^v-\alpha_i^v=2\pi/n_v,\,i=1,\ldots,n_v-1$. Finally, for an ordered set of points $\mathcal{S}^{v}=\{\mathcal{S}^{v}_{1},\ldots,\mathcal{S}^{v}_{n_v}\}\in \mathbb{R}^2$ we denote with $chull(\mathcal{S}^{v})$ their convex hull. Then, the proposed routine to build a convex under-approximation of the obstacle-free region 
starts with a regular polygon with $n_v-1$ faces inscribed in the ball $\mathcal{B}(x_{(1:2)}(k),d_{min}(k))=\{w\in\mathbb{R}^2:\|w-x_{(1:2)}(k)\|_2\leq d_{min}(k)\}$, and then expands each vertex $\mathcal{S}^{v}_{i},\,i=\,\ldots,n_v$, one at a time, as long as the convex hull of $\mathcal{S}^{v}$ does not contain any LiDAR reading. When the process is completed for all vertexes, the algorithm returns the wanted polytope in matrix inequality form as: 
\begin{equation}\label{eq:safe-set}
\mathcal{S}(k)\doteq\{x_{(1:2)}\in\mathbb{R}^2:A_c(k)x_{(1:2)} \leq b_c(k)\}
\end{equation}
To take into account the size of the drone, its maximum encumbrance is removed from the LiDAR readings. \noindent As an example, Fig. \ref{fig:build_poly} shows a few iterations of the proposed algorithm.
\begin{remark}\label{rem:strict_impl}
	By construction, the proposed algorithm 
	ensures that the current $(x_f,y_f)$ position of the drone,  $x_{(1:2)}(k)$, is always in the interior of $\mathcal{S}(k)$, i.e. in the safe set. 
	Moreover, this approach returns a valid convex under-approximation of the free space also when stopped at an intermediate step, which is an advantage when a strict real time implementation is needed.
\end{remark}

\subsection{Multi-trajectory model predictive control} \label{SS:mtMPC}
\begin{figure}[!htb]
 	\centering
 	\includegraphics[width=.7\columnwidth]{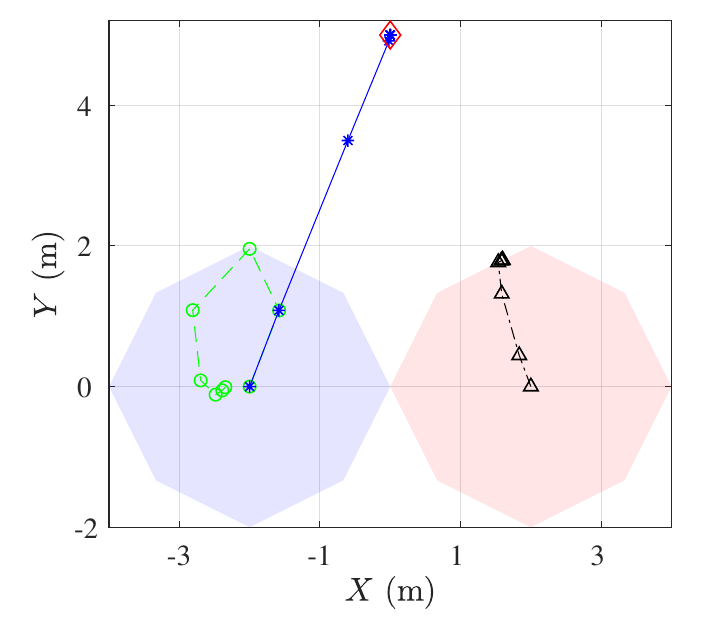}
 	\caption{Example illustrating the multi-trajectory concept (left) vs. a single-trajectory one (right). Blue line with `*': exploiting trajectory; green line with `$\circ$': safe trajectory; black dash-dotted line with `$\vartriangle$': single-trajectory approach. Red  `$\Diamond$': target. Colored polytopes represent the safe set $\mathcal{S}(k)$.\label{fig:motivation}}	
 \end{figure}

A common approach for autonomous vehicles to guarantee safety in a receding horizon framework is to consider a trajectory able to reach a stop inside $\mathcal{S}(k)$ within the prediction horizon $N\in\mathbb{N}$, see e.g. \cite{schouwenaars2004receding}.
 However, in our case such an approach can be too conservative, since the environment is unknown and the safe set $\mathcal{S}(k)$ is a convex under-approximation of the free space that generally changes over time. The proposed mt-MPC technique aims to reduce such conservativeness, by planning two trajectories: a ``safe'' one, fully contained in the safe set and reaching a steady state, and an ``exploiting'' one, which assumes that the current constraints are too conservative and is thus allowed to violate them. The two trajectories feature the same input at the current step and are allowed to diverge only afterwards in the prediction. To better illustrate this concept, let us consider the example in Fig. \ref{fig:motivation}, showing on the left the multi-trajectory approach and on the right a standard, single-trajectory one. 
 \\\noindent The shown trajectories are computed by solving a FHOCP using the control-oriented model \eqref{eq:control-oriented} and aiming to reach a target beyond the safe set. The single-trajectory approach forces the whole predicted trajectory to remain inside the safe set, minimizing the average distance from the target. This solution is optimal for the current safe set, but it does not consider a possible favorable evolution of the latter at the next time step. On the other hand, the multi-trajectory approach plans a much better (yet currently unfeasible) exploiting trajectory, but still retaining an alternative, guaranteed safe one in case the safe set does not expand towards the target. Since the approach is implemented in receding horizon, the  potential advantage is apparent by comparing the position reached at the next time step by the two approaches, see Fig. \ref{fig:motivation}. 
Let us denote with $x^I(j|k)$ the state of the exploiting trajectory of system \eqref{eq:control-oriented} at time $k+j$ predicted at time $k$, and with $x^{II}(j|k)$ that of the safe trajectory. 
We introduce the common input $\bm{u}^I(0|k)$ and the two input sequences $U^{I}=\left[\bm{u}^I(1|k)^{^T} \ \dots \ \bm{u}^I(N-1|k)^{^T}\right]^T$ 
and $U^{II}=\left[\bm{u}^{II}(1|k)^{^T} \ \dots \ \bm{u}^{II}(N-1|k)^{^T}\right]^T$, pertaining to the exploiting and safe trajectories. 
We consider the following cost function to track a reference $\bm{P}_t(k)$:
\begin{equation}
J(x(k),U,\bm{P}_t(k))= \sum_{j=1}^{N}\|x^I_{(1:3)}(j|k)-\bm{P}_t(k)\|_Q
\end{equation}
where $\|x\|_Q\doteq x^TQx$, $Q\in\mathbb{R}^{6\times6}$ is a symmetric positive-definite weighting matrix, and vector $U = [\bm{u}^I(0|k)^{^T} \ U^{I^T} \ U^{II^T}]^T$ $\in \mathbb{R}^{3(2N-1)}$. We are now in position to formulate the multi-trajectory FHOCP $\mathscr{P}(x(k),A_c(k),b_c(k),\bm{P}_t(k))$:\small
\begin{subequations}\label{eq:FHOCP}
\begin{gather}
	\label{eq:FHOCP-cost}
	\min\limits_U \;\;J(x(k),U,\bm{P}_t(k)) \\
	\text{subject to} \nonumber \\
	x^I(j|k) = Ax^I(j-1|k)+B\bm{u}^I(j-1|k),  \ j = 1,\dots,N\label{eq:constrX1dyn}\\	
	x^{II}(1|k) = x^I(1|k),\label{eq:constrX2dyn} \\
	x^{II}(j|k) = Ax^{II}(j-1|k)+B\bm{u}^{II}(j-1|k), \ j = 2,\dots,N\\
	-\bar{\bm{A}}\leq K_1\left(\bm{u}^I(0|k)-x^I_{(1:3)}(0|k)\right)-K_2x^I_{(4:6)}(0|k) \leq \bar{\bm{A}}\\
	-\bar{\bm{A}}\leq K_1\left(\bm{u}^{I,II}(j|k)-x^{I,II}_{(1:3)}(j|k)\right)-K_2x^{I,II}_{(4:6)}(j|k) \leq \bar{\bm{A}},\\
	\quad\quad\quad\quad\quad\quad\quad\quad\quad\quad\quad\quad\quad j = 1, \dots, N-1 \nonumber \\
	A_c(k)x^{I}_{(1:2)}(1|k)\leq b_c(k), \label{eq:constrX21}\\
	A_c(k)x^{II}_{(1:2)}(j|k) \leq b_c(k),  \ j = 2, \dots, N \label{eq:constrX22} \\
	-\bar{\bm{V}}\leq x^{I}_{(4:6)}(1|k) \leq \bar{\bm{V}},\\
-\bar{\bm{V}}\leq x^{I,II}_{(4:6)}(j|k) \leq \bar{\bm{V}}, \ j = 2, \dots, N \\
	 x^{II}_{(4:6)}(N|k) = 0^{3\times 1} \label{eq:termConstrX2}\\
 x^I(0|k) = x(k) \label{eq:constrX0}
\end{gather}
\end{subequations}\normalsize
where all equalities and inequalities are element-wise. The FHOCP \eqref{eq:FHOCP} is a convex QP that, if feasible, can be solved efficiently for a global minimizer. We denote its solution as $U^*(x(k),A_c(k),b_c(k),\bm{P}_t(k))$ and the corresponding optimal predicted state trajectories, with all elements stacked in single column vectors, as $X^{I*}(x(k),A_c(k),b_c(k),\bm{P}_t(k))\in\mathbb{R}^{6N}$ and $X^{II*}(x(k),A_c(k),b_c(k),\bm{P}_t(k))\in\mathbb{R}^{6N}$. Moreover, at any time $k$ let us denote with $\underline{k}(k)< k$ the latest sampling instant such that the FHOCP $\mathscr{P}(x(\underline{k}(k)),A_c(\underline{k}(k)),b_c(\underline{k}(k)))$ was feasible. 
%
The FHOCP (\ref{eq:FHOCP}) is embedded in the following receding horizon strategy:\\

\begin{figure}
	\centering
	\includegraphics[width=.78\columnwidth]{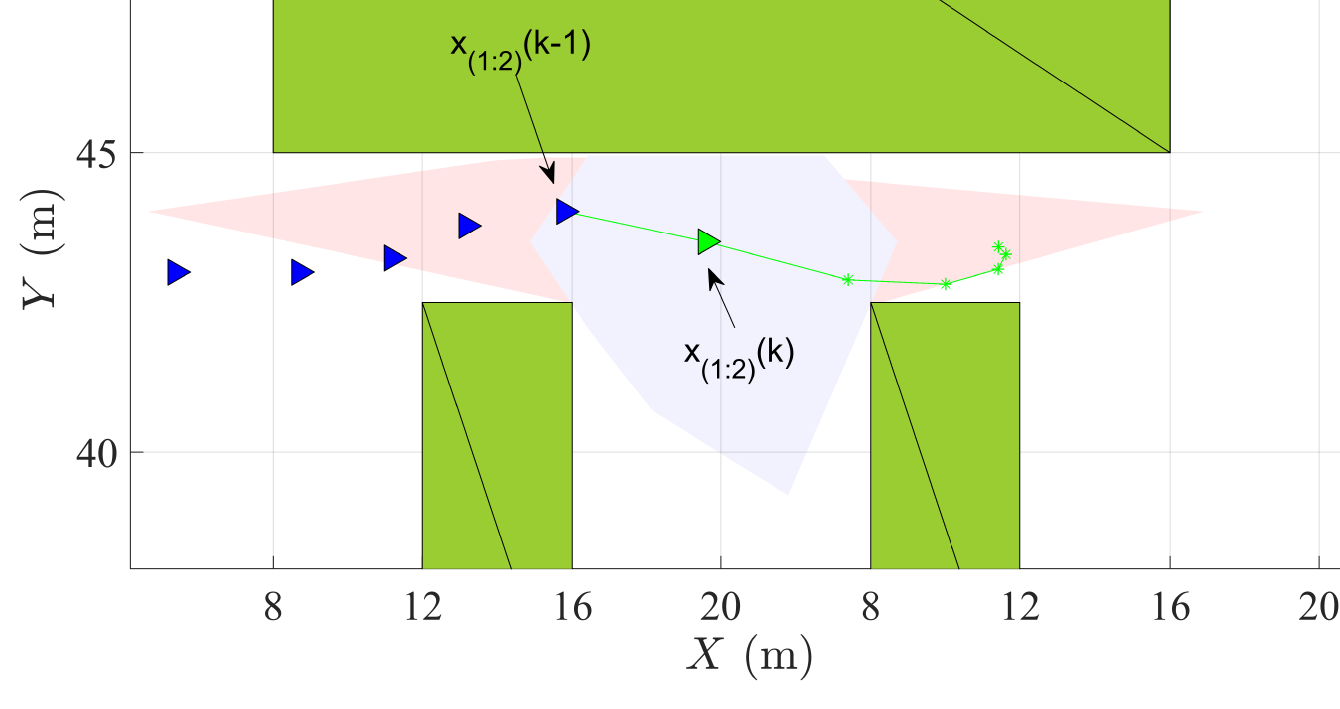}
	\caption{Illustration of the strategy adopted in Algorithm 1 to guarantee that a feasible FHOCP can be formulated at each $k$. The FHOCP that considers the safe set $\mathcal{S}(k)$ (blue) generated from $x_{(1:2)}(k)$ (green triangle) is unfeasible, but still the FHOCP that considers the safe set $\mathcal{S}(\underline{k}(k))$ (red) generated from $x_{(1:2)}(k-1)=x_{(1:2)}(\underline{k})$ is feasible.}	
	\label{fig:feasibility}
\end{figure}

\noindent\textbf{Algorithm 1}: multi-trajectory MPC (mt-MPC)
\begin{itemize}
\item[\textbf{1)}] At time $k$ collect the LiDAR measurements and compute the matrices $A_c(k)$, $b_c(k)$.
\item[\textbf{2)}] 
\begin{itemize}
\item[\textbf{if\;\;}] $\mathscr{P}(x(k),A_c(k),b_c(k),\bm{P}_t(k))$ is feasible, apply to the system the first control input in the optimal sequence $U^*(x(k),A_c(k),b_c(k),\bm{P}_t(k))$. Set $\underline{k}(k+1)=k$ and store $A_c(k)$, $b_c(k)$ as $A_c(\underline{k}(k+1))$, $b_c(\underline{k}(k+1))$.
\item[\textbf{else}] solve $\mathscr{P}(x(k),A_c(\underline{k}(k)),b_c(\underline{k}(k)),\bm{P}_t(k))$ and apply to the system the first control input in the optimal sequence $U^*(x(k),A_c(\underline{k}(k)),b_c(\underline{k}(k)),\bm{P}_t(k))$. Set $\underline{k}(k+1)=\underline{k}(k)$.
\end{itemize}
\item[\textbf{3)}]  set $k=k+1$ and go to 1).
\end{itemize} 
Since for a time-invariant environment the safe set generated 
depends only on $x(k)$, the mt-MPC approach results in a dynamic, state-feedback control law with internal state $\underline{k}(k)$ and input $\bm{P}_t(k)$:
\begin{equation}\label{eq:mt-MPC-law}
	\begin{array}{rcl}
	 \underline{k}(k+1)&=&\eta(x(k),\underline{k}(k))\\
	 \bm{u}(k)&=&\kappa(x(k),\underline{k}(k),\bm{P}_t(k))
	 \end{array}
\end{equation}
where functions $\eta:\mathbb{R}^6\times\mathbb{Z}\rightarrow\mathbb{Z}$ and $\kappa:\mathbb{R}^6\times\mathbb{Z}\times\mathbb{R}^3\rightarrow\mathbb{R}^6$ are implicitly defined by Algorithm 1. The resulting close loop system is
\begin{equation}\label{eq:mt-MPC-cl}
\begin{array}{rcl}
\underline{k}(k+1)&=&\eta(x(k),\underline{k}(k))\\
x(k+1)&=&Ax(k)+B\kappa(x(k),\underline{k}(k),\bm{P}_t(k))
\end{array}	
\end{equation}
In Algorithm 1, the role of variable $\underline{k}(k)$ and of the corresponding set $\mathcal{S}(\underline{k}(k))$ (i.e. of matrices $A_c(\underline{k}(k)),b_c(\underline{k}(k))$) is to guarantee that at each time step a feasible FHOCP can be formulated, notwithstanding the time-varying nature of the safe convex set $\mathcal{S}(k)$.  This is illustrated in Fig. \ref{fig:feasibility}. Such a guarantee holds however only when unknown but static obstacles are considered, as in our problem. In presence of time-varying obstacles, additional assumptions and different approaches would be required, currently subject of our research.
\noindent In turn, the capability of formulating a feasible FHOCP at each $k$ implies that the mt-MPC approach achieves an obstacle-free trajectory, as shown by the following result.
\begin{lemma}
	\label{lemma_rec_feas}
	Assume the FHOCP (\ref{eq:FHOCP}) at time $k = 0$ is feasible and $x(0)\notin\mathcal{O}$, i.e. the drone is initially in the obstacle-free region. Then, the trajectory of the close loop system \eqref{eq:mt-MPC-cl} is such that $x(k)\notin\mathcal{O},\;\forall k>0$.
\end{lemma}
\begin{proof}
	At $k=0$, problem $\mathscr{P}(x(0),A_c(0),b_c(0),\bm{P}_t(0))$ is solved and $\underline{k}(1)$ is set equal to 0. For any $k\geq0$, let us denote with $\tilde{U}^{II*}(k)=[\tilde{\bm{u}}^{II*}(1|k)^T,\ldots,\tilde{\bm{u}}^{II*}(N-1|k)^T]$ the optimal safe input sequence computed by the mt-MPC algorithm, be it by solving $\mathscr{P}(x(k),A_c(k),b_c(k),\bm{P}_t(k))$ or $\mathscr{P}(x(k),A_c(\underline{k}(k)),b_c(\underline{k}(k)),\bm{P}_t(k))$ (see Algorithm 1), and with $\tilde{x}^{II*}(k+N|k)$ the corresponding safe terminal state.\\
	\noindent Then, at each $k>0$, there are only two possibilities.\\
	\noindent \textbf{A)} If $\mathscr{P}(x(k),A_c(k),b_c(k),\bm{P}_t(k))$ is feasible,  then at time $k+1$ we have
	$x(k+1)=X^{II*}_{(1:2)}(x(k),A_c(k),b_c(k),\bm{P}_t(k))\in\mathcal{S}(k)$, see \eqref{eq:constrX1dyn}-\eqref{eq:constrX2dyn} and \eqref{eq:constrX21}.\\
	\noindent \textbf{B)} Conversely, if $\mathscr{P}(x(k),A_c(k),b_c(k),\bm{P}_t(k))$ is not feasible, problem $\mathscr{P}(x(k),A_c(\underline{k}(k)),b_c(\underline{k}(k)),\bm{P}_t(k))$ is solved, for which a feasible sequence can be built as $[\tilde{U}^{II*}(k-1)^T \tilde{x}^{II*}_{1:3}(k+N|k)^T]^T$. In fact, constraints \eqref{eq:termConstrX2} impose that $\tilde{x}^{II*}(k+N|k)$ is a steady state (zero velocity) and the corresponding steady input is $\bm{u}=\bm{P}_{ref}=\tilde{x}^{II*}_{(1:3)}(k+N|k)$ (compare eq. \eqref{eq:acceleration}), which also satisfies acceleration and velocity constraints. Thus, in this case we have $x(k+1)=X^{II*}_{(1:2)}(x(k),A_c(\underline{k}(k)),b_c(\underline{k}(k)),\bm{P}_t(k))\in\mathcal{S}(\underline{k}(k))$.\\
	\noindent Therefore, we have that, in both cases \textbf{A)} and \textbf{B)}, $x(k+1)$ belongs to a set $\mathcal{S}(j)$, with $j\leq k$. Now, by construction, 
	whenever $x(j)\notin \mathcal{O}$ then the corresponding set $\mathcal{S}(j)$ is an under-approximation of the obstacle-free region, i.e. $\mathcal{S}(j)\cap O_i=\emptyset,\,\forall i=1,\ldots,N_o$, meaning that in both cases \textbf{ A)} and \textbf{B)} we have $x(k+1)\notin \mathcal{O}$. We thus demonstrated that $x(k)\notin \mathcal{O}\Rightarrow x(k+1)\notin \mathcal{O}$. The result is then proven by induction, considering that $x(0)\notin\mathcal{O}$ by assumption.
\end{proof}
\begin{remark}
	According to $Lemma$ \ref{lemma_rec_feas}, the proposed approach guarantees an obstacle free-trajectory for the control-oriented model \eqref{eq:control-oriented}. The latter however exhibits a mismatch with respect to the real-system, in our case represented by the nonlinear model of the drone 
	. Such an error is indeed rather small, as it can be appreciated in Fig. \ref{fig:ControlOrientModel} (and a similar situation typically occurs with a real-world, position-controlled drone). The model-plant mismatch can be estimated and bounded by suitable uncertainty quantification techniques (see, e.g. \cite{milanese2013bounding,lauricella2020set}), and then accounted for by the MPC law using robust formulations \cite{bemporad1999robust,MaSR05,limon2010robust}. The use of these approaches is subject of our future research, beyond the scope of this paper. However, a remarkable aspect is that in the randomly generated simulations presented in Section \ref{S:SimRes}, where we employed the nonlinear system to simulate the real-world drone, we never observed any collision with the obstacles.
\end{remark}
\subsection{Temporary target shifting strategy}\label{SS:temp-target}
\begin{figure}[!hbt]
	\centering
	\includegraphics[width=.75\columnwidth]{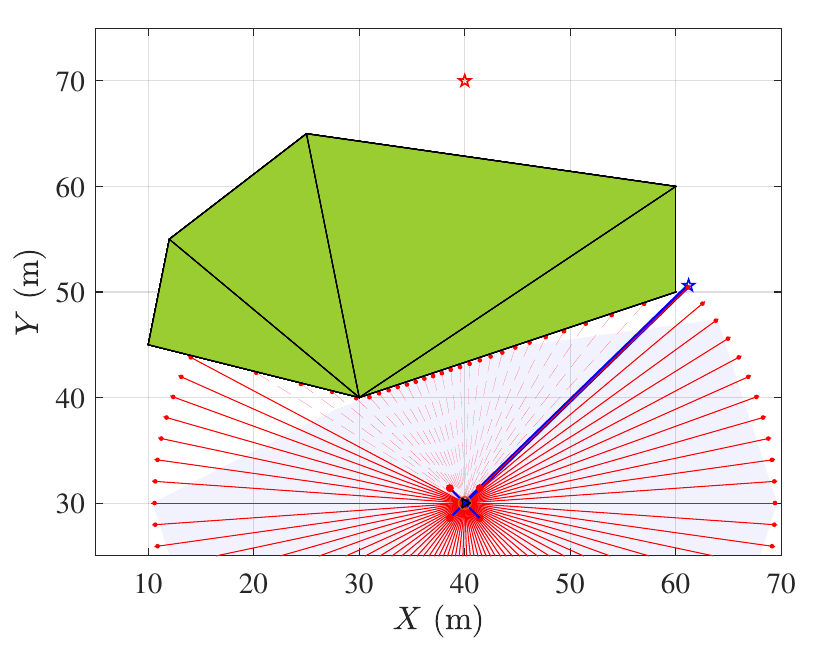}
	\caption{Example of temporary target shifting. Blue star: $\hat{\bm{P}}_{t}(k)$; red star: $\bm{P}_t(k)$; red dashed lines: LiDAR measurements with index $i\notin \mathcal{I}(k)$, red solid lines: LiDAR measurements with index $i\in \mathcal{I}(k)$.}	
	\label{fig:temp_targ}
\end{figure}

A common problem in optimization-based autonomous navigation with obstacles is the possibility that the system is not able to circumnavigate an obstacle in between the current drone position and the target, as shown for example in Fig. \ref{fig:temp_targ}.
In such a situation, if no countermeasure is taken, the drone can simply stop at a locally optimal position, meaning that any allowed movement temporarily implies a growth of distance from the target. To avoid this situation, we propose a strategy named temporary target shifting. Let $v_t=\bm{P}_t(k)-x_{(1:2)}(k)$ be the vector connecting the drone to the target.
When the LiDAR readings whose angular position is closest to that of direction $v_t$ do not detect any obstacle (up to the maximum range $R_L$), the original target $\bm{P}_t(k)$ in used. Otherwise, a temporary target $\hat{\bm{P}}_t(k)$ is chosen as follows.
Consider the set of indexes corresponding to the LiDAR readings reporting the maximum distance:
\begin{equation}
	\mathcal{I}(k)=\left\{\bar{j}:\bar{j}=\arg\max_{i=0,\dots,M-1}\rho_i(k)\right\}
\end{equation}
Then, the temporary target is obtained as the LiDAR reading, among those with index $\bar{j}\in \mathcal{I}$, that is closest to the target: 
\begin{equation}\label{eq:temp-target}
 \hat{\bm{P}}_{t}(k) = \min\limits_{\bar{j}\in \mathcal{I}}{\left\|s_{j}(k) -v_t\right\|_2}
\end{equation}
This strategy is illustrated in  Fig. \ref{fig:temp_targ} as well. The temporary target $\hat{\bm{P}}_{t}(k)$ is then held constant until the drone reaches it, or until the target direction becomes obstacle-free again, whichever condition happens first. This choice avoids situations where the drone starts moving back and forth behind an obstacle because the temporary target is periodically switched between the two visible edges. If more than one solution to \eqref{eq:temp-target} exists, we take the one with smallest index $\bar{j}$.
\section{Simulation results}\label{S:SimRes}
\begin{figure}[!htb]
	\centering
	\includegraphics[width=.79\columnwidth]{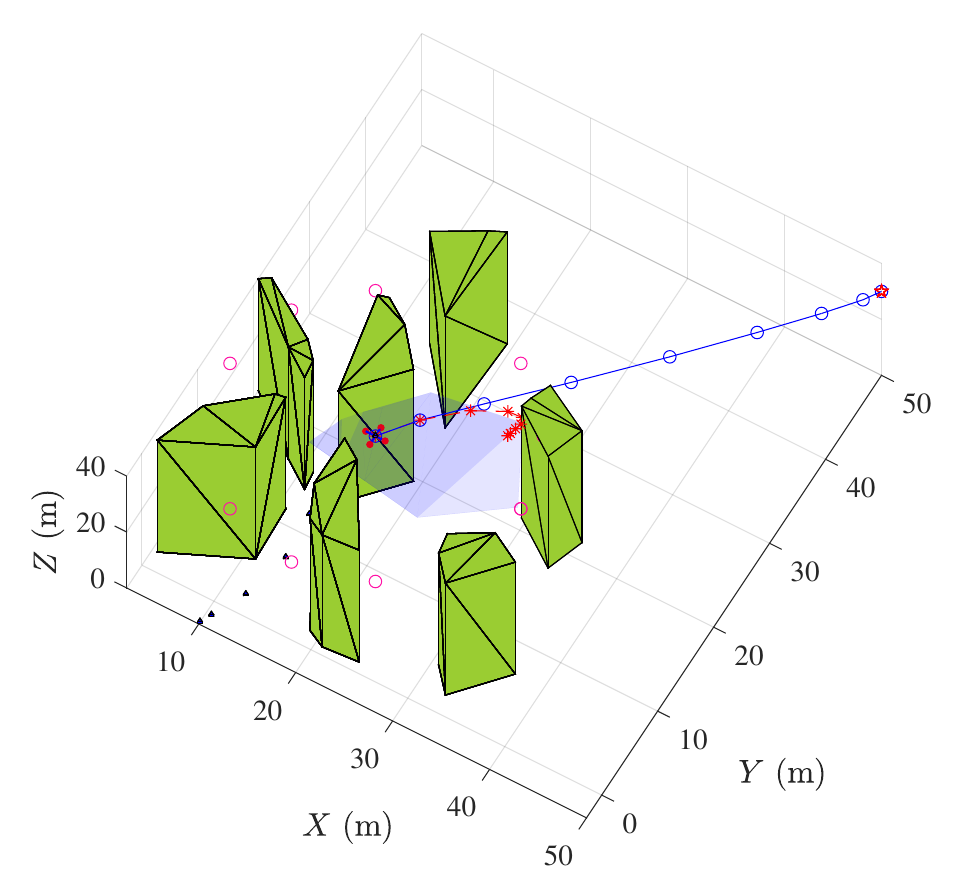}
	\caption{Simulation of the mt-MPC approach. Black triangles: past drone positions. Light blue polytope: set $\mathcal{S}(k)$. Pink circles: LiDAR's maximum field of view. Dashed red line: safe trajectory. Solid blue line: exploiting trajectory. Red star: target $\bm{P}_t$. Obstacles beyond the LiDAR's field of view are not shown. 
	}	
	\label{fig:MPC}
\end{figure}

\begin{figure}
	\centering
	\includegraphics[width=.8\columnwidth]{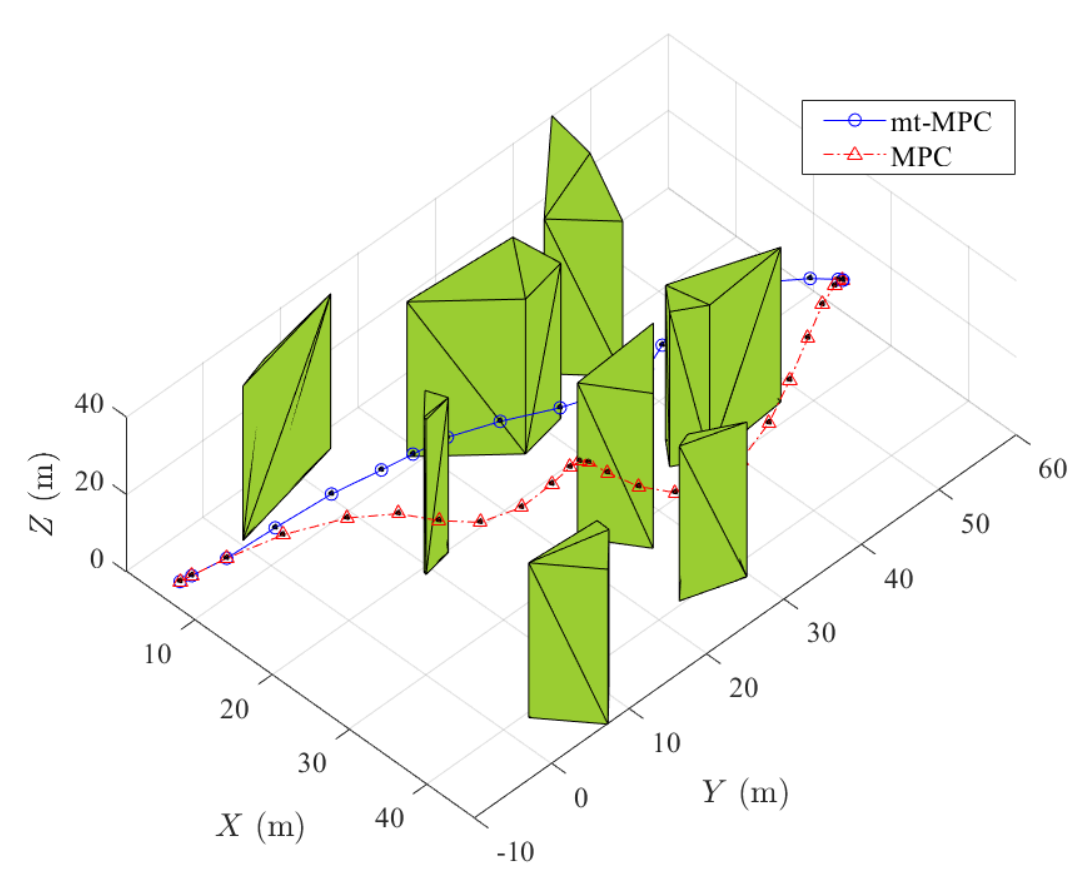}
	\caption{Simulation results. Trajectories obtained with mt-MPC (solid blue line) and MPC (dash-dotted red line).}	
	\label{fig:comp}
\end{figure}

We evaluated the effectiveness of the proposed approach via numerical simulations with the layout reported in Fig. \ref{fig:hierarchical}, i.e. using the actual nonlinear model as ``real system''. 
We employed a Quad-Core Intel Core i7 (3.6 GHz, 32 GB) on MATLAB 2020a 
and the QPs are solved using Yalmip \cite{yalmip}. 
Fig. \ref{fig:MPC} shows the UAV during navigation in a typical simulation test: 
the convex under-approximation $\mathcal{S}(k)$, the ``safe'' trajectory and the ``exploiting'' one can be easily distinguished.
The approach exhibits good performance, driving quickly the drone to its target without collisions. A movie of the  simulation example shown in Fig. \ref{fig:MPC} is available at \cite{YoutubeVideo}.
On average, in our tests the numerical solution of the QP \eqref{eq:FHOCP} required about $40\,$ms per sampling step, and about $60\,$ms in addition to compute the under-approximation of the obstacle-free region. 
\noindent We compared the mt-MPC approach to a standard MPC, which still employs  Algorithms 1, however with a single-trajectory in the FHOCP, subject to all the operational and safety constraints. We used the same tuning parameters in the cost function for the two approaches. 
Fig. \ref{fig:comp} shows a comparison between the resulting close-loop trajectories. Both drive the UAV to its target without collisions, however obtaining very different paths. For a more thorough comparison, we ran a series of $N_{sim} = 100$ problems with randomly generated initial state, obstacles, and target, but all presenting a similar layout
, i.e. where the drone has to traverse an area with many unknown obstacles. In each test $i$, we measured the time $T_i$ needed to reach the target, as well as the cumulative close loop tracking error $J_i=\sum\limits_{k=0}^{T_i}||\bm{P}(k)-\bm{P}_{t_i}||^2_2$.
Then, we considered the following average quantities as performance indicators: $\bar{J} = \dfrac{1}{N_{sim}}\sum\limits_{i=1}^{N_{sim}}J_i$, $\bar{T}=\sum_{i=1}^{N_{sim}}T_i$.
Finally, we compared also the computational effort required to solve the FHOCP in the two cases.
The obtained results indicate that the mt-MPC improves both the average tracking error ($-6.6\%$) and the average time to reach the target ($-6.8 \%$) with respect to the standard MPC technique, however with a higher computational time  ($+12.6\%$ to solve the QP).
The presented results thus confirm that the approach has good potential in terms of performance improvement, with a rather small increase in computational effort.
\section{Conclusions}
A novel MPC formulation, named mt-MPC, has been presented, where multiple trajectories are considered in the same optimization problem to trade-off conflicting objectives, in this case safety and exploitation. The technique has been applied to the navigation of an UAV in a-priori unknown environment. 
A novel approach to approximate the feasible set with a convex polytope exploiting only real-time LiDAR measurements has been used, together with a strategy to ensure recursive feasibility and guarantee obstacle avoidance.
mt-MPC showed a good performance improvement with respect to standard MPC, at the cost of a limited increase in computational time. Current research is aimed to apply the mt-MPC concept to nonlinear and/or reconfigurable systems, and to include also a learning component in the problem, pertaining to the environment (exploration/mapping) and to the autonomous system (uncertainty and model mismatch quantification).
\bibliographystyle{IEEEtran}
\bibliography{IEEEabrv,BiblioMultiTraj}  
\end{document}